\documentstyle[prl,aps]{revtex}
\input epsf

\tighten

\begin{document}


\newcommand{\Bd}{{\dot B}}
\newcommand{\Cd}{{\dot C}}
\newcommand{\fd}{{\dot f}}
\newcommand{\hd}{{\dot h}}
\newcommand{\ep}{\epsilon}
\newcommand{\vp}{\varphi}
\newcommand{\al}{\alpha}
\newcommand{\be}{\begin{equation}}
\newcommand{\ee}{\end{equation}}
\newcommand{\bea}{\begin{eqnarray}}
\newcommand{\eea}{\end{eqnarray}}
\def\gapp{\mathrel{\raise.3ex\hbox{$>$}\mkern-14mu
              \lower0.6ex\hbox{$\sim$}}}
\def\gsim{\gapp}
\def\lapp{\mathrel{\raise.3ex\hbox{$<$}\mkern-14mu
              \lower0.6ex\hbox{$\sim$}}}
\def\lsim{\lapp}
\newcommand{\PSbox}[3]{\mbox{\rule{0in}{#3}\includegraphics{#1}\hspace{#2}}}
\def\Tr{\mathop{\rm Tr}\nolimits}
\def\su#1{{\rm SU}(#1)}

\title{Causality and Cosmic Inflation}

\author{
Tanmay Vachaspati\footnote[1]{\tt txv7@po.cwru.edu.}
and
Mark Trodden\footnote[2]{\tt trodden@theory1.phys.cwru.edu}}
\address
{Department of Physics,
Case Western Reserve University,
10900 Euclid Avenue,
Cleveland, OH 44106-7079, USA.}

\wideabs{
\maketitle

\begin{abstract}
\widetext
In the context of inflationary models with a pre-inflationary
stage, in which the Einstein equations are obeyed, the weak
energy condition is satisfied, and spacetime topology is trivial,
we argue that homogeneity on super-Hubble scales must be
assumed as an initial condition.
Models in which inflation arises from field dynamics in a 
Friedman-Robertson-Walker background fall into this class but 
models in which inflation originates at the Planck epoch,
{\it eg.} chaotic inflation, may evade this conclusion.
Our arguments rest on causality and general relativistic 
constraints on the structure of spacetime.
We discuss modifications to existing scenarios that may
avoid the need for initial large-scale homogeneity.

\end{abstract}
\pacs{}
}

\narrowtext

It is well recognized that an early inflationary epoch 
can explain several of the observed features of the 
present universe \cite{Gut81,Lin90}. The remarkable homogeneity of the
universe as measured by COBE, the flatness of the 
universe indicated by some of the current cosmic data, 
the distribution of structure, and the absence of magnetic
monopoles may all be simultaneously explained by invoking about 
60 e-folds of cosmic inflation. This remarkable fact has spurred
considerable effort in building models that realize an inflationary
phase of the universe. The goal of inflationary models is to
explain how an assumed non-inflationary universe after the big bang 
develops into an inflationary universe at some epoch. Eventually,
after some 60 e-folds,
the universe must gracefully exit the inflationary stage and enter
the radiation epoch of standard cosmology. 

There exist alternative explanations for
some of the cosmological observations that inflation
so naturally explains.
The distribution of structure may follow from topological
defects \cite{VilShe94}; the absence of magnetic monopoles 
from details of particle physics \cite{LanPi},
or the interaction of domain walls and magnetic monopoles
\cite{DvaLiuVac97}. Other observed features of the universe 
are harder to explain by non-inflationary means. If the universe
is indeed flat, it would be hard to explain this observation
without invoking inflation.  (It is known that certain inflationary 
models can lead to a non-flat universe, and so flatness is not a 
generic prediction of inflation but one of certain models.) 
Finally, the homogeneity of the universe is virtually impossible
to explain without invoking inflation and this is a key 
compelling feature of the theory.

The ability of inflation to smooth out the universe on superhorizon
scales is an effective mechanism to explain the observed
homogeneity only if the inflation itself does not require violations
of causality. This means that we must assume a pre-inflationary
epoch of the universe from which a small patch of the universe underwent
inflation entirely by causal processes. Note that causality dictates
that the inflation must be ``local''. In other words, any spacelike
section of the boundary of the inflating region must not extend beyond 
the causal horizon of the pre-inflationary spacetime.

The question we address here is: under what conditions is it possible to 
have {\em local} inflation?

The embedding of an inflating region 
(not necessarily undergoing {\em exponential} inflation) 
within an exterior cosmology 
is constrained by the nature of matter in the universe. This is
best seen by employing the Raychaudhuri equation for the divergence
of a congruence of future directed, affinely parametrized null 
geodesics. This congruence is taken to be normal to
a two dimensional sphere centered at the origin of coordinates
and may be in- or out-going 
(i.e. directed towards or away from the origin of coordinates, 
respectively).
Let us denote the tangent vector field to the congruence 
by $N^a$. Then the divergence $\theta$ is defined by
\begin{equation}
\theta = \nabla_a N^a \ .
\label{theta}
\end{equation}
The Raychaudhuri equation is:
\begin{equation}
{{d\theta} \over {d\tau}} + {1\over 2} \theta^2 =
-\sigma_{ab}\sigma^{ab} + \omega_{ab}\omega^{ab}
-R_{ab}N^a N^b
\label{raychaudhari}
\end{equation}
where $\tau$ is the affine parameter, $\sigma_{ab}$
is the shear tensor, $\omega_{ab}$ the twist tensor
and $R_{ab}$ the Ricci tensor. 
(We follow the conventions of Wald \cite{Wald}.)
The shear tensor is purely spatial and hence its contribution to 
the right-hand side is positive. The twist tensor vanishes since
the congruence of null rays is 
taken to be hypersurface orthogonal. Then, 
\begin{equation}
{{d\theta} \over {d\tau}} + {1\over 2} \theta^2 \leq -R_{ab}N^a N^b 
\label{rayineq}
\end{equation}
If Einstein's equations hold then
\begin{equation}
R_{ab}N^a N^b = 8 \pi T_{ab}N^a N^b \ ,
\label{einsteineq}
\end{equation}
and if the weak energy condition is satisfied 
($T_{ab} \xi^a \xi^b \geq 0$ for any timelike vector $\xi^a$),
then by continuity
\begin{equation}
T_{ab}N^a N^b \geq 0 \ .
\label{wec}
\end{equation}
Putting these conditions together we obtain
\begin{equation}
{{d\theta} \over {d\tau}} + {1\over 2} \theta^2 \leq 0 \ .
\label{condition}
\end{equation}
For our purposes, however, it proves sufficient to use the weaker
condition
\begin{equation}
{{d\theta} \over {d\tau}} \leq 0 \ .
\label{usedcondition}
\end{equation}

Regions of 
a spherically symmetric 
spacetime in which the divergences of both in- and 
out-going rays, 
normal to spatial two dimensional spheres 
centered at the origin, are negative (positive) will be
referred to as trapped (antitrapped) regions. 
Regions in which in-going
rays have negative divergence (that is, are converging)
but out-going rays have positive divergence will be called
``normal'', since this is the behaviour in flat spacetime.
Then the condition~(\ref{usedcondition}) says that
a converging null geodesic cannot start to diverge prior to
having reached the origin of coordinates, or {\it focussed}. 
In other words, in-going null rays cannot 
start out in normal regions and then enter an antitrapped
region. This becomes the constraint in patching together
an inflationary region in a background cosmology.

\begin{figure}[tbp]
\caption{\label{local}
A Penrose diagram for local inflation. The arrow denotes a 
future directed, radial, affinely parametrized null geodesic 
from the exterior spacetime into the inflating region.
Shaded regions are antitrapped, unshaded regions are normal.
}
\vskip 0.5 truecm
\epsfxsize = \hsize \epsfbox{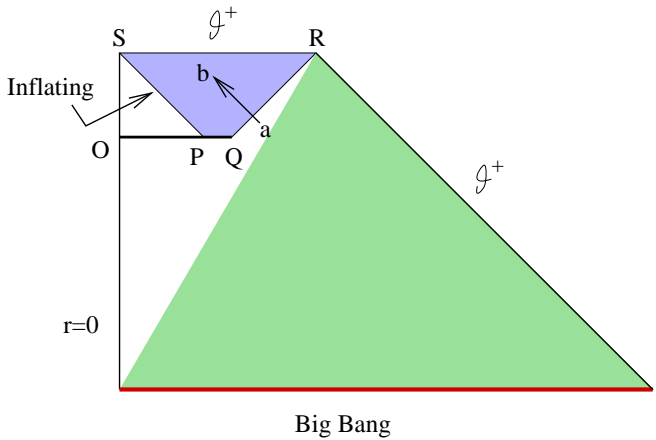}
\end{figure}

Consider a topologically trivial universe such as shown in
Fig. \ref{local}. The universe starts out in a big bang and contains
a normal region and an antitrapped region at distances larger
than some distance that depends on the details of the cosmology.
Now consider a patch of this region that starts to inflate.
The patch is denoted by the horizontal line OQ,
and has a physical size that we will denote by $x_Q$. 
The section OP of the line OQ denotes a spatial patch
equal to the size of the inflationary horizon $H_{inf}^{-1}$.
For inflation to occur, one assumes that vacuum energy must 
dominate over a region larger than the inflationary horizon 
distance, and so
\begin{equation}
x_Q \geq x_P = H_{inf}^{-1} \ .
\label{xphinf}
\end{equation}
Further, a straightforward calculation for ingoing null
rays in spacetimes with the metric of the inflating region given by 
a flat FRW metric

\begin{equation}
ds^2 = -dt^2 + a^2 (t) [ dr^2 + r^2 d\Omega^2 ] \ ,
\end{equation}
yields
\begin{equation}
\theta = {2 \over {a(t)}} \left ( H - {1\over x} \right ) \ ,
\label{thetafrw}
\end{equation}
where, as usual, $H = {\dot a}/a$, and
\begin{equation}
x = a(t) r
\label{xdefn}
\end{equation}
where $r$ is the coordinate of a null ray. Eq. (\ref{thetafrw}) 
shows that the region with physical distance larger than 
$H_{inf}^{-1}$ in the inflating region is antitrapped. 
Hence the region PQRS in Fig. \ref{local} is antitrapped. 
The crucial question is: what are the
allowed positions of the point P?

\begin{figure}[tbp]
\caption{\label{global}
A Penrose diagram for local inflation in which ingoing null
geodesics that enter the inflating region
emanate from antitrapped regions.
}
\vskip 0.5 truecm
\epsfxsize = \hsize \epsfbox{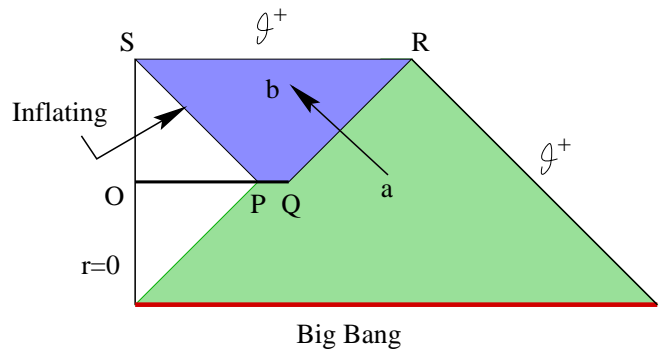}
\end{figure}

In Fig. \ref{local} we show the situation where P is not located in 
the antitrapped region of the background cosmology. Then light rays
such as shown in Fig. \ref{local} (from point $a$ to point $b$)
can enter the inflating antitrapped 
region from the external normal region (unshaded in the figure).
While the ingoing rays are in the normal region, $\theta$ is
negative, but once they enter the antitrapped inflationary region
$\theta$ must become positive. This is forbidden by the condition
in eq. (\ref{usedcondition}). Hence, we must conclude that the
outer boundary PQR of the antitrapped inflating region must lie 
entirely inside the antitrapped region of the background cosmology as
shown in Fig. \ref{global}. This is the key constraint on inflating
spacetimes derived in this paper 
and, except for spherical symmetry,
is independent of the background cosmology. 

To appreciate the constraint, it is useful to think of the
situation when the background cosmology is a flat
Friedman-Robertson-Walker universe with a scale factor $a(t)$.
Then, the boundary of the antitrapped region of the background
universe is given by
\begin{equation}
x_{FRW} (t) = H_{FRW}^{-1} (t) \ .
\label{xhfrw}
\end{equation}
Now, since the point P must lie within the background antitrapped
region
\begin{equation}
x_P \geq x_{FRW} (t_P ) \ ,
\label{xpxfrw}
\end{equation}
which yields
\begin{equation}
x_Q \geq x_P = H_{inf}^{-1} \geq H_{FRW}^{-1} (t_P) \ .
\label{hdshfrw}
\end{equation}
This says that the size of the initial inflationary patch
must be greater than the
inflationary horizon, which must be larger than the
background FRW inverse Hubble size
at the time inflation starts. That is, the
conditions appropriate for inflation to occur must be satisfied
over a patch that is larger than the FRW 
inverse Hubble scale\footnote{Based on homogeneous FRW cosmologies, 
a similar conclusion (${\dot H}_{FRW} \le 0$) was reached by Linde 
and collaborators \cite{Lin90Sec8.6,LinLinSec5,LinLinMez}. Though 
leading to the same qualitative conclusion, there are differences
in the final results since Linde and collaborators compare the
Hubble scales for homogeneous universes at two different times
while we are considering inhomogeneous universes and the 
Hubble scales are compared at the same time. However, as
we also describe in this paper, 
the conclusion has been used as an argument against the
viability of new inflationary models by these authors.} 

Note that the inverse Hubble distance, $H^{-1}$, 
can be different from the causal horizon (the distance light
has propagated from the big bang). However, $H^{-1}$ is in 
most cases still a large patch compared to length scales
over which particle physics processes occur that can homogenize 
the universe.
Also, for a flat, radiation dominated FRW cosmology, $H^{-1}$ 
coincides with the causal horizon. We conclude that inflationary 
model building must assume homogeneity on super-Hubble 
scales. In this sense, inflationary models that attempt to 
obtain inflation within a background FRW universe cannot 
explain the homogeneity of the observed universe.

Our result is consistent with
the result due to Farhi and Guth 
\cite{FarhiGuth} who found that it is impossible to create
an inflationary universe in the laboratory subject to the
Einstein equations, the weak energy condition and the absence
of singularities. On small enough scales in an expanding universe,
it should be possible to ignore the background expansion and
then the Farhi-Guth result should be applicable.
This is consistent with our result since we find that the
Hubble scale of the background spacetime provides a lower bound 
on the size of the inflating patch. 
If one admits the possibility of inflating false vacuum
bubbles born at a 
singularity, the spacetime diagrams drawn by Blau, Guendelman
and Guth \cite{BlaGunGut} show that the inflating region
emerges from a white hole interior in which all two spheres
are antitrapped. Then, once again, the boundary of the inflating
region borders an antitrapped region.

In \cite{GolPir}, Goldwirth and Piran 
numerically solved the Einstein equations together with a scalar field
and found that inflation is obtained only if homogeneous
initial conditions are assumed over a length scale that
encompasses several horizons. (Similar numerical analyses were 
also performed by Kung and Brandenberger \cite{KunBra}.) 
Our result generalizes and proves this numerical finding. 

It is also worth pointing out that chaotic inflation 
\cite{Lin90} does not fall within the purview of our result since, 
in this model, inflation starts at the Planck epoch with homogeneity 
assumed on the Planck scale. 

Hence we conclude that, within the conditions described above,
local inflation is not possible. However, observations
indicate homogeneity of the early universe on super-horizon scales
and this needs an explanation. It is indeed possible that the
initial homogeneity required by inflation 
- or even the homogeneity of the entire visible universe -
occurred just by chance. 
Whether this is a satisfactory resolution of the observed homogeneity 
of the universe is largely a matter of personal taste and possibly 
anthropic considerations. 

\begin{figure}[tbp]
\vskip 0.5 truecm
\epsfxsize = \hsize \epsfbox{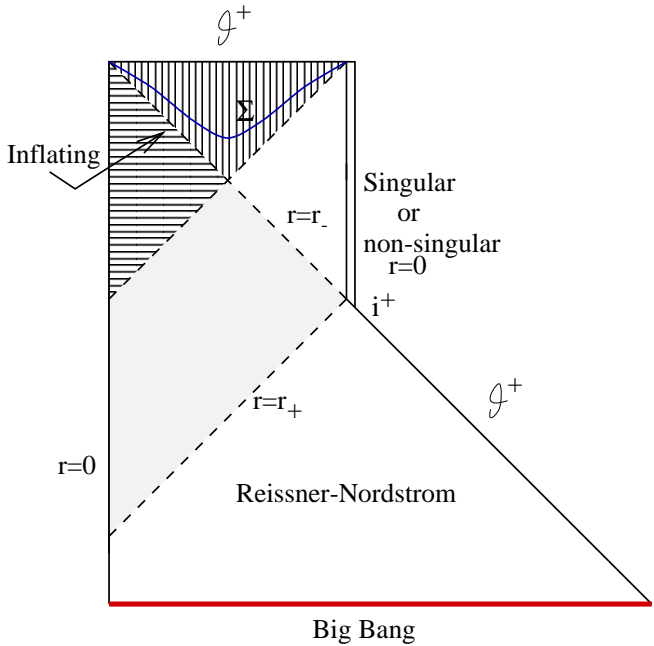}
\vskip 0.5 truecm
\caption{\label{topinf}
A Penrose diagram for local inflation as in topological inflation
with magnetic monopoles. Initial data must be provided on a
spacelike hypersurface $\Sigma$ entirely within the inflating region.
}
\end{figure}

Let us now discuss 
the conditions under which local inflation can occur without
assuming accidental homogeneity on large scales. 
The first possibility is that the weak energy condition (for diagonal, 
spherically symmetric fluid energy-momentum tensors, this amounts to 
assuming $\rho \geq 0$ and $\rho +P \geq 0$.) may be 
violated. For this one would need exotic forms of matter in
the early universe. An attractive alternative is that
quantum effects could give rise to effective violations of the
weak energy condition. (However these violations
are constrained by the Ford-Roman inequalities \cite{ForRom}.)
Whether quantum effects can be sufficient
to lead to local inflation is an interesting question that has not
yet been answered (an early related attempt was made 
in \cite{FarGutGuv}). 
The second possibility is that the Einstein
equations may be modified, leading to changes in
eqs. (\ref{einsteineq}) and (\ref{condition}). This is
possible, for example, if we have a non-minimally coupled scalar field
in the model. To us, this way out seems to be the best
possibility especially in view of modern particle theories
in which such scalar fields are abundant. A third possibility
may be to have a topologically non-trivial background
universe. 
Such universes have attracted significant attention recently \cite{glenn}
and should be investigated further. A fourth possibility
is the one that occurs in topological inflation within
magnetic monopoles as we discussed in detail
in \cite{BorTroVac98}. Here the inflation is manifestly
local and causal but is preceded by a singularity or
topology change (see Fig. \ref{topinf}). Conflict with 
our constraint is 
avoided because there are no null rays that enter the inflating 
region from the external region. The singularity or topology
change plays the role of a mini big bang for the inflating 
spacetime, though is somewhat different in character from an
FRW big bang since it is timelike. In any case, predictability
in the inflating universe is lost because of signals that can
emanate from the singularity or topology changing event.
It is not possible to evolve to the
inflating region from data on a spacelike hypersurface in the
pre-inflationary epoch. Instead, initial data must be provided
on a spacelike surface ($\Sigma$) within the inflating region. 

To summarize, we have argued that 
inflationary models
based on the classical Einstein equations, the weak energy conditions, 
and trivial topology, require initial homogeneity on super-Hubble
scales. 
Inflation with no requirements of initial large-scale homogeneity
can be achieved with one or more of the following conditions:
1) violations of the classical
Einstein equations, say due to non-minimally coupled scalar fields,
2) violations of the weak energy condition in the early universe,
3) non-trivial topology of the universe,
4) the birth of the universe directly into an inflating universe,
that is, the absence of a pre-inflationary epoch, such as might 
occur in specific inflationary models, {\it eg.} chaotic inflation,
and/or in the context of quantum cosmology \cite{Vil82,HarHaw83,Linqc} 

We would like to thank Arvind Borde for guidance, 
Alan Guth and Andrei Linde for extensive discussions,
Jaume Garriga, Lawrence Krauss, Glenn Starkman, Alex Vilenkin 
and Bob Wald for comments.
This work was supported by the Department of Energy (D.O.E.).

\end{document}